\documentclass[%
 reprint,
 amsmath,amssymb,
 aps,
]{revtex4-2}

\usepackage{float}
\usepackage{graphicx}
\usepackage{dcolumn}
\usepackage{bm}

\usepackage{hyperref}

\begin{document}

\title{Defect-Defect Interactions in the Buckling of Imperfect Spherical Shells}

\author{Fani Derveni}%
\affiliation{%
 Ecole Polytechnique Fédérale de Lausanne (EPFL)\\
    Flexible Structures Laboratory\\
    CH-1015 Lausanne, Switzerland
}%

\author{Arefeh Abbasi}
\affiliation{%
 Ecole Polytechnique Fédérale de Lausanne (EPFL)\\
    Flexible Structures Laboratory\\
    CH-1015 Lausanne, Switzerland
}

\author{Pedro M. Reis}%
 \email{pedro.reis@epfl.ch}
\affiliation{%
 Ecole Polytechnique Fédérale de Lausanne (EPFL)\\
    Flexible Structures Laboratory\\
    CH-1015 Lausanne, Switzerland
}%


\begin{abstract}
We perform finite element simulations to study the impact of defect-defect interactions on the pressure-induced buckling of thin, elastic, spherical shells containing two dimpled imperfections. Throughout, we quantify the critical buckling pressure of these shells using their knockdown factor. We examine cases featuring either identical or different geometric defects and systematically explore the parameter space, including the angular separation between the defects, their widths and amplitudes, and the radius-to-thickness ratio of the shell. As the angular separation between the defects is increased, the buckling strength initially decreases, then increases before reaching a plateau.
Our primary finding is that the onset of defect-defect interactions, as quantified by a characteristic length scale associated with the onset of the plateau, is set by the critical buckling wavelength reported in the classic shell-buckling literature. Beyond this threshold, within the plateau regime, the buckling behavior of the shell is dictated by the largest defect.
\end{abstract}

\maketitle
\noindent \textbf{Dedication:} We dedicate this manuscript to Prof. Kyung-Suk Kim, a truly inspiring scholar in our Mechanics community and a beacon of inspiration, rigor, creativity, and intellectual generosity. Prof. Kim's mastery of opening new research directions and revisiting classic problems, always with fresh eyes, have been a constant source of inspiration for us. The corresponding author is especially grateful to Prof. Kim for the exceptional support, guidance, and mentoring he received over the years.

\section{Introduction}
\label{sec:introduction}
The buckling of elastic shell structures is highly sensitive to imperfections~\cite{von1939buckling,von1940influence,hutchinson1970postbuckling}; a problem that is relevant across length scales, from viruses~\cite{lidmar2003virus} and colloidal capsules~\cite{datta_delayed_2012} to large storage tanks~\cite{godoy_buckling_2016}.
Even if this is a long-standing classic subject~\cite{Zoelly1915,tsien_theory_1942, koiter_over_1945, kaplan_nonlinear_1954,bijlaard1960elastic, Seide1960,kiernan_elastic_1963,Carlson1967,kobayashi_influence_1968,hutchinson_effect_1971,budiansky1972buckling,babcock_shell_1983}, the past decade has seen a revival in the study of the buckling of shells and their imperfection sensitivity~\cite{gerasimidis2023foreword}. For a historical perspective and a more thorough contextualization of the modern account of single-defect shell buckling, we direct the reader to Refs.~\cite{hutchinson2016buckling,hutchinson_john_w._nonlinear_2017, thompson_probing_2017,virot_stability_2017,jimenez_technical_2017,gerasimidis_establishing_2018,marthelot_buckling_2017,lee2019evolution,yan2020buckling,yadav2021nondestructive,abbasi2021probing,yan2021magneto,pezzulla2022geometrically,abbasi2023comparing,gerasimidis2023foreword}. 

The canonical question, which remains challenging despite decades of research, is: \textit{What are the critical conditions for the buckling of an imperfect shell?} Recently, in an effort to address this question, an experimental technique has been developed for fabricating spherical shells containing a single dimpled imperfection, which can be engineered precisely~\cite{lee2016fabrication}. Subsequent buckling studies utilizing this model system demonstrated that if the geometry of the imperfection is characterized in detail, the critical pressure can be predicted accurately, either using the Finite Element Method (FEM) or via numerical solutions of the shell-theory equations~\cite{lee2016geometric}. The knockdown factor, defined as the ratio between the critical buckling pressure of the imperfect shell and that of the equivalent perfect one\cite{Zoelly1915}, is the commonly used metric in these studies. For realistic shells, predicting the knockdown factor, which is always less than unity, is notoriously challenging.

Beyond the model system of a single-defect shell, we have recently investigated the more realistic case of a large number of geometric imperfections distributed randomly over the surface of a spherical shell~\cite{derveni2023probabilistic}. Importantly, we evidenced that given an input log-normal distribution for the amplitude of defects, the resulting knockdown factor is described by a 3-parameter Weibull distribution, a  finding that places shell buckling in the broader class of extreme-value statistics phenomena \cite{fisher1928limiting,weibull1939phenomenon,weibull1951statistical,jayatilaka_statistical_1977,bavzant2009scaling,le2015modeling}. In that study, we also found that interactions between two adjacent defects, depending on the defect-to-defect separation, can potentially strengthen or weaken the shell in comparison to the single-defect case. There is a similar problem for cylindrical shells, both with a single defect \cite{hutchinson1965buckling,kyriakides1992bifurcation,virot_stability_2017,gerasimidis_establishing_2018,yadav2021nondestructive,groh2023probing} or distribution of defects~\cite{amazigo1969buckling,elishakoff1982reliability,elishakoff1985reliability,ELISHAKOFF201235}. Even though there have been some studies on the buckling of cylindrical shells containing two defects~\cite{wullschleger2006numerical,fan2019critical}, to the best of our knowledge, a systematic exploration of defect-defect interactions in the buckling of \textit{spherical} shells has not been tackled to date.

Here, we study the buckling of imperfect hemispherical shells containing two dimpled defects. The geometric properties of these two imperfections can be either identical or different. Methodologically, we conduct FEM simulations, which have been previously validated thoroughly against experiments~\cite{derveni2023probabilistic}. First, we focus on how the angular separation between the two defects affects the knockdown factor, characterizing how the interaction regime is impacted by the width and amplitude of the imperfections.
Then, we compare the threshold of the defect-defect separation for the onset of interactions to the theoretical prediction of the full wavelength of the classic critical buckling wavelength for a spherical shell~\cite{Hutchinson1967}.
Our main finding is that the arc length associated with the defect-defect interaction threshold depends directly on the radius-to-thickness ratio of the shell, scaling linearly with this critical buckling wavelength.

Our paper is organized as follows. First, in Sec.~\ref{sec:probdefinition}, we define the problem at hand and outline the research questions. Next, in Sec.~\ref{sec:FEM}, we describe the FEM simulations employed in our study. In Sec.~\ref{sec:hypothesis}, we present a first set of results on the influence of the radius-to-thickness ratio on the buckling behavior of shells containing two defects. More detailed results for shells with identical defects are provided in Sec.~\ref{sec:identical} and with different defects in Sec.~\ref{sec:different}. Finally, in Sec.~\ref{sec:conclusions}, we summarize the conclusions of our study and offer suggestions for future research directions.
\section{Problem Definition}
\label{sec:probdefinition}

\begin{figure}[t]
    \centering
    \includegraphics[width=\columnwidth]{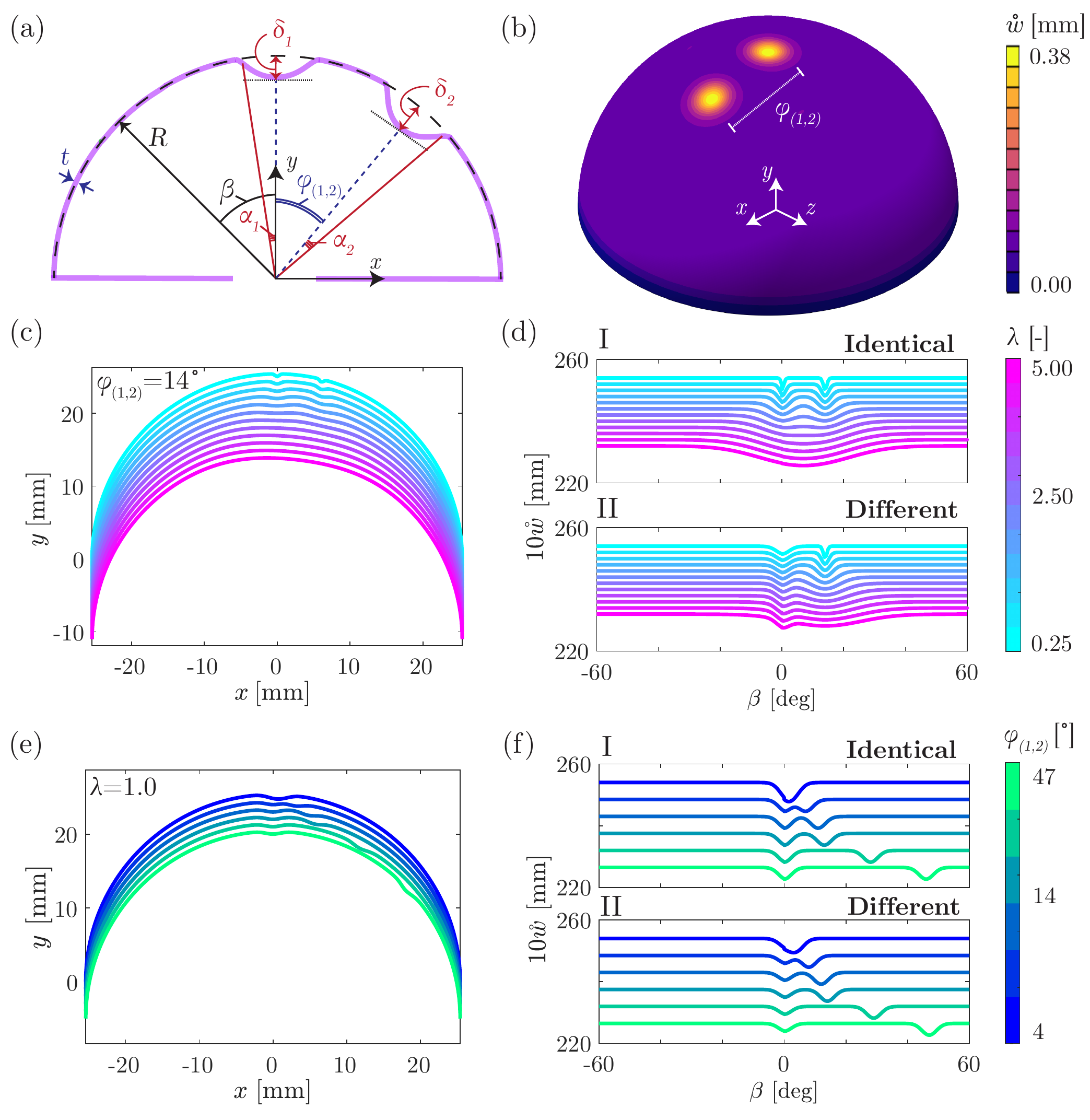}
    \caption{Reference geometry of the imperfect hemispherical shell with two dimpled defects. (a) 2D schematic, defining all relevant geometric quantities. (b) 3D representation; the shade (see colorbar) represents the radial deviation $\mathring{w}$ from a perfect sphere. (c,e) Geometric profiles of identical-defect shells for (c) fixed $\overline{\delta}=1.5$, $\varphi_{(1,2)}=14^\circ$ and varying $\lambda_i$, and (e) fixed $\overline{\delta}=1.5$, $\lambda_i=1.0$ and varying $\varphi_{(1,2)}$. (d,f) Radial deflection, $\mathring{w}$, versus zenith angle, $\beta$, for (d) constant $\varphi_{(1,2)}=14^\circ$ between (dI) identical defects  with various $\lambda_i$ or (dII) different defects  with various $\lambda_2$. (f) Similar data, with constant $\lambda_i=1$, for (fI) identical defects with various $\varphi_{(1,2)}$ or (fII) different defects with various $\varphi_{(1,2)}$. The representative cases for identical defects (dI, fI) have $\overline{\delta}_i=1.5$, and the different-defects cases (dII,fII) have $\overline{\delta}_1=1$, $\overline{\delta}_2=1.5$ and $\lambda_1=1$.
    For clarity, all profiles are offset in panels (c,d) by $1$~mm, in (e) by $2$~mm, and in (f) by $5.5$~mm downwards. Also, the $\mathring{w}$ profiles in panels (d) are shown with an amplification factor of $10$.}
    \label{fig:geometry}
\end{figure}

We consider a thin, elastic, and hemispherical shell of radius, $R$, and thickness, $t$, as illustrated in Fig.~\ref{fig:geometry}(a,b). The shell is clamped at the equator and contains \textit{two} geometric imperfections. In their undeformed configuration, each defect is shaped as a Gaussian dimple, with the following radial deviation from the perfect spherical geometry:
\begin{align}\label{eqn:geom_gaussiandimple}
\mathring{w}_i(\alpha)= -\delta_i e^{-(\alpha/\alpha_i)^2}, 
\end{align}
where the indices $i=\{1,2\}$ represent each of the two defects, $\alpha$ is the local angular distance corresponding to each defect (measured from their centers), $\alpha_i$ is the half-angular width of the $i$th defect, and $\delta_i$ is its amplitude (maximum radial deviation of the mid-surface of the shell). The global angular (zenith) coordinate, $\beta$, is defined from the pole ($\beta=0$), where the first defect ($i=1$) is always located. The other defect is at $\beta_2$. Following conventional practice in shell-buckling studies~\cite{kaplan_nonlinear_1954,koga1969axisymmetric},
the defect amplitude of each defect is normalized as $\overline{\delta}_i=\delta_i/t$, while the width is normalized as
$\lambda_i=[12(1-\nu^2)]^{1/4}\,(R/t)^{1/2}\,\alpha_i$. Here, $\nu$ is the Poisson's ratio of the material. The shell thickness, $t$, is kept constant throughout so that we focus only on geometric imperfections, unlike previous work on through-thickness defects~\cite{yan2020buckling} or elasto-plastic dents~\cite{gerasimidis2021dent}. 

First, we will analyze shells containing two identical defects: $\lambda=\lambda_1=\lambda_2$ and $\overline{\delta}=\overline{\delta}_1=\overline{\delta}_2$. Subsequently, we will  consider the scenario of two different defects; $\lambda_1\neq\lambda_2$ and/or $\overline{\delta}_1\neq\overline{\delta}_2$. Since the $i=1$ defect is always positioned at the shell pole ($\beta=0$) and the $i=2$ defect is at $\beta_2$, the angular separation (center-to-center) between the two defects is $\varphi_{(1,2)}=\beta_2$. To facilitate the discussion on defect-defect interactions later in the manuscript, it is important to define an alternative angular separation:
\begin{equation}\label{eq:phi_star_definition}
\varphi_{(1,2)}^*=\varphi_{(1,2)}-m\frac{\alpha_1+\alpha_2}{\sqrt2},
\end{equation}
where $m=\{1,\,2,\,3\}$ is an integer. The different values of $m$ correspond to successively excluding  wider portions from the core of the defects when considering their angular separation. A more comprehensive discussion on this point will be provided in  Sec.~\ref{sec:identical}. Finally, recalling Eq.~\eqref{eqn:geom_gaussiandimple}, the combined profile of a shell with two dimples is
\begin{equation}\label{eqn:2defect_profile}
\mathring{w}(\beta,\,\theta) =  \mathring{w}_1(0,\,0)+\mathring{w}_2(\varphi_{(1,2)},\,\theta_2),
\end{equation}
where $\beta$ and $\theta$ are the \textit{global} zenith and azimuthal spherical (polar) coordinates, respectively.

Figs.~\ref{fig:geometry}(c-f) depict representative examples of the mid-surface profile of a shell with $R/t=100$. These profiles are visualized within the great plane that intersects the shell and passes through the centers of the two imperfections. Note that, given the localized (dimpled) profile in Eq.~(\ref{eqn:2defect_profile}), the shells are \textit{not} axisymmetric, and the profiles shown in Fig.~\ref{fig:geometry} are solely for illustration purposes. Figs.~\ref{fig:geometry}(c,e) show the Cartesian profiles in the $y$-$x$ great plane; for clarity, all profiles are offset vertically (see caption for details). As an alternative representation, the $\mathring{w}(\beta)$ curves in Figs.~\ref{fig:geometry}(d,f) correspond to the radial deviation from a perfect hemisphere as a function of the global zenith angle,  $\beta\in[-60,\,60]^\circ$. These limiting angles are chosen as the maximum location of the defects to avoid interactions with the equator boundary~\cite{derveni2023probabilistic}.
When their widths, $\lambda_i$, are too large (Figs.~\ref{fig:geometry}c,d) or when their angular separation, $\varphi_{(1,2)}$, is too small (Figs.~\ref{fig:geometry}e,f), the two defects can merge to form a single defect.

Following a similar approach as in previous studies~\cite{lee2016geometric,marthelot_buckling_2017,yan2020buckling,abbasi2021probing,derveni2023probabilistic}, we depressurize the clamped hemispherical shell until buckling occurs. Given the actual critical buckling pressure of the imperfect shell, $p_\mathrm{max}$, the knockdown factor is defined as $\kappa=p_\mathrm{max}/p_\mathrm{c}$,
where $p_\mathrm{c}$ is the classic prediction for the respective perfect shell geometry~\cite{Zoelly1915, lee2016geometric}. Our goal is to characterize how $\kappa$ for a shell with the two-defect geometry specified above depends on the
following geometric parameters: $\overline{\delta}_i$, $\lambda_i$, $\varphi_{(1,2)}$, and $R/t$. 
We will give particular attention to identifying the regimes where the interactions between the two defects induce non-trivial changes in $\kappa$.

Our main contribution will be the definition of a threshold arc length for the separation between the two defects,
beyond which their interactions become negligible. We will consider two versions of this separation-arclength threshold: $l_{p}=R\varphi_{p(1,2)}$, defined from center-to-center of the defect, and $l_{p}^*=R\varphi_{p(1,2)}^*$, adjusted to account for edge effects of the defects using
$\varphi_{(1,2)}^*$ introduced in Eq.~(\ref{eq:phi_star_definition}). We provide evidence that this latter arclength, with $m=1$, is set by
\begin{equation}
    l_{p}^*\approx l_\mathrm{c}=2\pi[12(1-\nu^2)]^{-1/4}\sqrt{Rt},
    \label{eq:criticalbucklingwavelength}
\end{equation}
where $l_\mathrm{c}$, computed in the seminal work by Hutchinson~\cite{Hutchinson1967}, is the theoretical critical buckling wavelength for a spherical shell. More technically, $l_\mathrm{c}$ is the full wavelength of the axisymmetric bifurcation mode at the equator of the shell. 

In our previous work~\cite{derveni2023probabilistic}, we presented preliminary evidence for the result in Eq.~(\ref{eq:criticalbucklingwavelength}), but only with a single value of $R/t=110$. Hence, we were unable to fully test Eq.~(\ref{eq:criticalbucklingwavelength}). In the present study, we will change this radius-to-thickness ratio within the range $R/t\in[100,\, 500]$ to examine how $l_{p}^*$ relates to $ l_\mathrm{c}$. Furthermore, in Ref.~\cite{derveni2023probabilistic}, we reported evidence for the potential interactions between nearby defects and how they can lead to stronger or weaker shells in comparison to single-defect shells. However, the data in that study was limited to a few specific cases. In the present work, we will explore the various geometric parameters of the system  systematically and seek to characterize how defect-defect interactions impact $\kappa$ for spherical shells containing two imperfections.

\section{Methodology: FEM Simulations}
\label{sec:FEM}

We performed full 3D simulations using the Finite Element Method (FEM) with the commercial software ABAQUS/Standard~\cite{Abaqus:2014}. In our prior work~\cite{derveni2023probabilistic,abbasi2023comparing}, we validated this approach against precision experiments similar to the multi-defects geometry considered here. 
Each quarter of the hemispherical shell is discretized in the meridional and azimuthal using four-noded S4R shell elements: a total of 67500 elements for shells with $R/t\leq300$ and 187500 elements for shells with $R/t\geq400$.
This level of discretization was deemed suitable after conducting a thorough mesh-convergence analysis.
To set the initial geometry of the imperfect shell, we initiated with a perfect hemispherical mesh. Subsequently, we introduced nodal displacements according to the desired profiles of the two imperfections, following Eq.~(\ref{eqn:2defect_profile}), with varying values for the geometric parameters ($\overline{\delta}_i$, $\lambda_i$, $\varphi_{(1,2)}$). The shell thickness remained constant throughout the simulations.

The shells were subject to uniform live pressure on their outer surface, while their equator was set as a clamped boundary. We employed a Riks (static) solver with the following parameters for the shells with $R/t\leq300$: an initial arc length increment of $0.1$, a minimum increment of $10^{-5}$,  and a maximum increment of $0.5$. For the thinnest shells with $R/t\geq400$, the corresponding parameters of the Riks solver were $0.002$, $10^{-10}$, and $0.2$, respectively. Geometric nonlinearities were considered throughout the analysis. 

The hemispherical shells were modeled using the material properties of vinylpolysiloxane (VPS-32, Elite Double 32, Zhermack) as a neo-Hookean and incompressible solid; the material had a Poisson's ratio of $\nu\approx0.5$ and Young's modulus of $E = 1.26 \,$MPa. These material properties were chosen to match those of previous shell-buckling experiments~\cite{lee2016geometric,marthelot_buckling_2017,yan2020buckling,abbasi2021probing,derveni2023probabilistic} used to validate our FEM-simulation approach. The geometric parameters of the two-defect imperfect shells were varied in the following ranges: $\overline{\delta}_i\in[0.5,\,3]$, $\lambda_i\in[0.25,\,5]$, $R/t\in[100,\,500]$ (constant $R=25.4\,\mathrm{mm}$, varying $t$) and $\varphi_{(1,2)}\in[1,\,60]^\circ$. 

\section{Hypothesis for the defect-defect interaction regime}
\label{sec:hypothesis}

We start our investigation by quantifying how the knockdown factor, $\kappa$, of the two-defects shells depends on the radius-to-thickness ratio, $R/t$. Throughout, we will focus on numerical experiments conducted using the FEM simulation approach described in the preceding section.

\begin{figure}[h!]
    \centering
    \includegraphics[width=0.82\columnwidth]{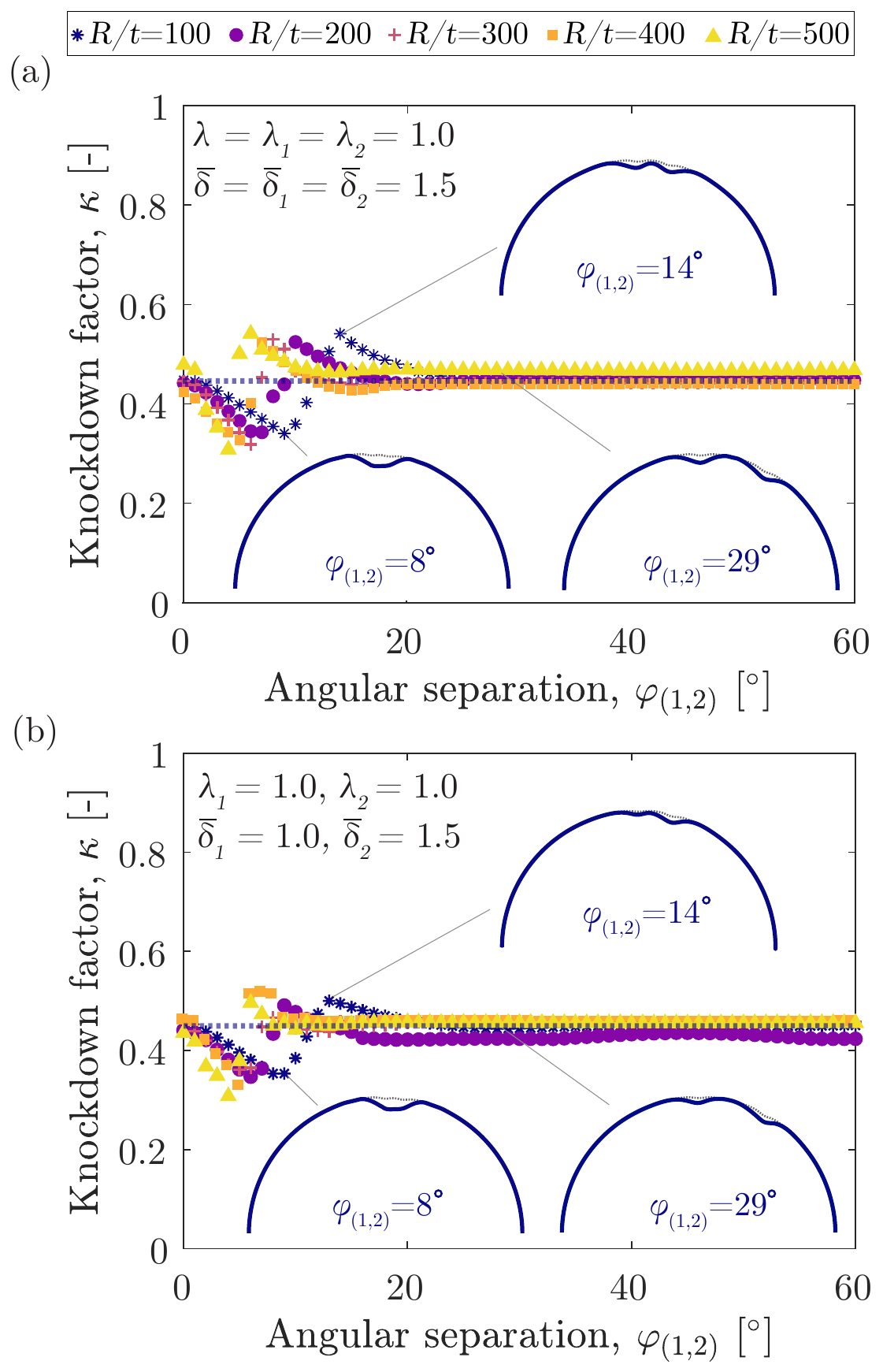}
    \caption{Knockdown factor, $\kappa$, as a function of angular separation, $\varphi_{(1,2)}$, for (a) identical and (b) different defects. The respective values of $\lambda_i$ and $\overline{\delta}_i$ are provided in the legend of each plot. Shells with varying radius-to-thickness ratio, $R/t$, are considered, as indicated in the top legend (common to both panels).
    Insets: Greater-plane profiles of imperfect shells with $R/t=100$ and different values of $\varphi_{(1,2)}$ in their original configurations (dotted lines) and at the onset of buckling (solid lines). The radial deviation of the latter is amplified by a factor of 3 for visualization purposes.  The horizontal dashed lines correspond to the $\kappa$ values of a single-defect shell with $R/t=100$ and $(\overline{\delta},\,\lambda)=(1.5,\,1)$.}
    \label{fig:kappa_vs_Rovert}
\end{figure}

In Fig.~\ref{fig:kappa_vs_Rovert}, we plot $\kappa$ versus the defect-defect angular separation, $\varphi_{(1,2)}$, for shells comprising either (a) two identical or (b) two different defects, at several values of $R/t$. For now, we set the amplitudes and widths of the defects as follows. For the case of identical defects (Fig.~\ref{fig:kappa_vs_Rovert}a), we fixed $\overline{\delta}=1.5$ and $\lambda=1$. For the case of different defects  (Fig.~\ref{fig:kappa_vs_Rovert}b), we fixed $\overline{\delta}_1=1$, $\overline{\delta}_2=1.5$ and $\lambda_1=\lambda_2=1$.
All curves are non-monotonic as a function of $\varphi_{(1,2)}$: $\kappa$ first decreases, reaching a minimum ($\kappa_\mathrm{min}$), then increases to a maximum ($\kappa_\mathrm{max}$), and subsequently decreases to a constant plateau value ($\kappa_\mathrm{p}$). As suggested in Ref.~\cite{derveni2023probabilistic}, this non-monotonic behavior at small values of $\varphi_{(1,2)}$ arises from defect-defect interactions. By contrast, in the plateau region at large values of $\varphi_{(1,2)}$, the largest defect dominates. Note that the horizontal dashed lines in Fig.~\ref{fig:kappa_vs_Rovert} correspond to $\kappa$ values for a single-defect shell with $(\overline{\delta},\,\lambda)=(1.5,\,1)$ and $R/t=100$, aligning with the plateaus of all the two-defects curves. The identical-defects shells (Fig.~\ref{fig:kappa_vs_Rovert}a) exhibit higher values of $\kappa_\mathrm{max}$ than the different-defects shells (Fig.~\ref{fig:kappa_vs_Rovert}b), suggesting that defect-defect interactions are less pronounced in the latter case.

To help visualize the buckling process, the insets of Fig.~\ref{fig:kappa_vs_Rovert} offer representative snapshots of the greater-plane (2D) profiles obtained from the FEM simulations for shells with $R/t=100$ and various defect-defect angular separations. Near $\kappa_\mathrm{min}$ (e.g., $\varphi_{(1,2)}=8^\circ$), the two defects are almost superimposed, resulting in a reduced knockdown factor (cf. Eq.~\ref{eqn:2defect_profile}). For intermediate separations (e.g., $\varphi_{(1,2)}=14^\circ$), near $\kappa_\mathrm{max}$, the region between the two defects acts as a constraint for buckling, leading to higher values of $\kappa$. When the two defects are sufficiently far apart (e.g., $\varphi_{(1,2)}=29^\circ$), in the plateau region, the largest defect dominates the buckling. 

All the plotted data sets in Fig.~\ref{fig:kappa_vs_Rovert}, with varying $R/t$ values, exhibit the aforementioned non-monotonic behavior of $\kappa(\varphi_{(1,2)})$.  However, as $R/t$ increases, the interaction regions (before the plateau is reached) progressively shift to lower values of $\varphi_{(1,2)}$. This observation highlights the influence of the radius and thickness of the shell on the defect-defect interactions. We hypothesize that the threshold angular separation, below which defects interact and above which the plateau begins, is directly related to $\sqrt{Rt}$; the characteristic length scale associated with the balance between bending and stretching effects~\cite{abbasi2021probing}. Consequently, we anticipate that the onset of the plateau in the $\kappa(\varphi_{(1,2)})$ curves is directly related to the critical buckling wavelength, $l_c~\sim \sqrt{Rt}$, as expressed in Eq.~\eqref{eq:criticalbucklingwavelength}. Without wanting to spoil a surprise, the results in the next section will confirm this hypothesis. 

\section{Interactions between two identical defects}
\label{sec:identical}

In this section, we focus solely on imperfect shells with two \textit{identical} defects. The  angular separation between their centers, $\varphi_{(1,2)}$, can be recast as the defect-defect separation \textit{arc length}, $l=R\varphi_{(1,2)}$.
Our objective is to quantify the dependence of the FEM-computed knockdown factor, $\kappa$, for these shells on $l$, $R/t$, $\overline{\delta}$, and $\lambda$.

\begin{figure}[h!]
    \centering
    \includegraphics[width=1\columnwidth]{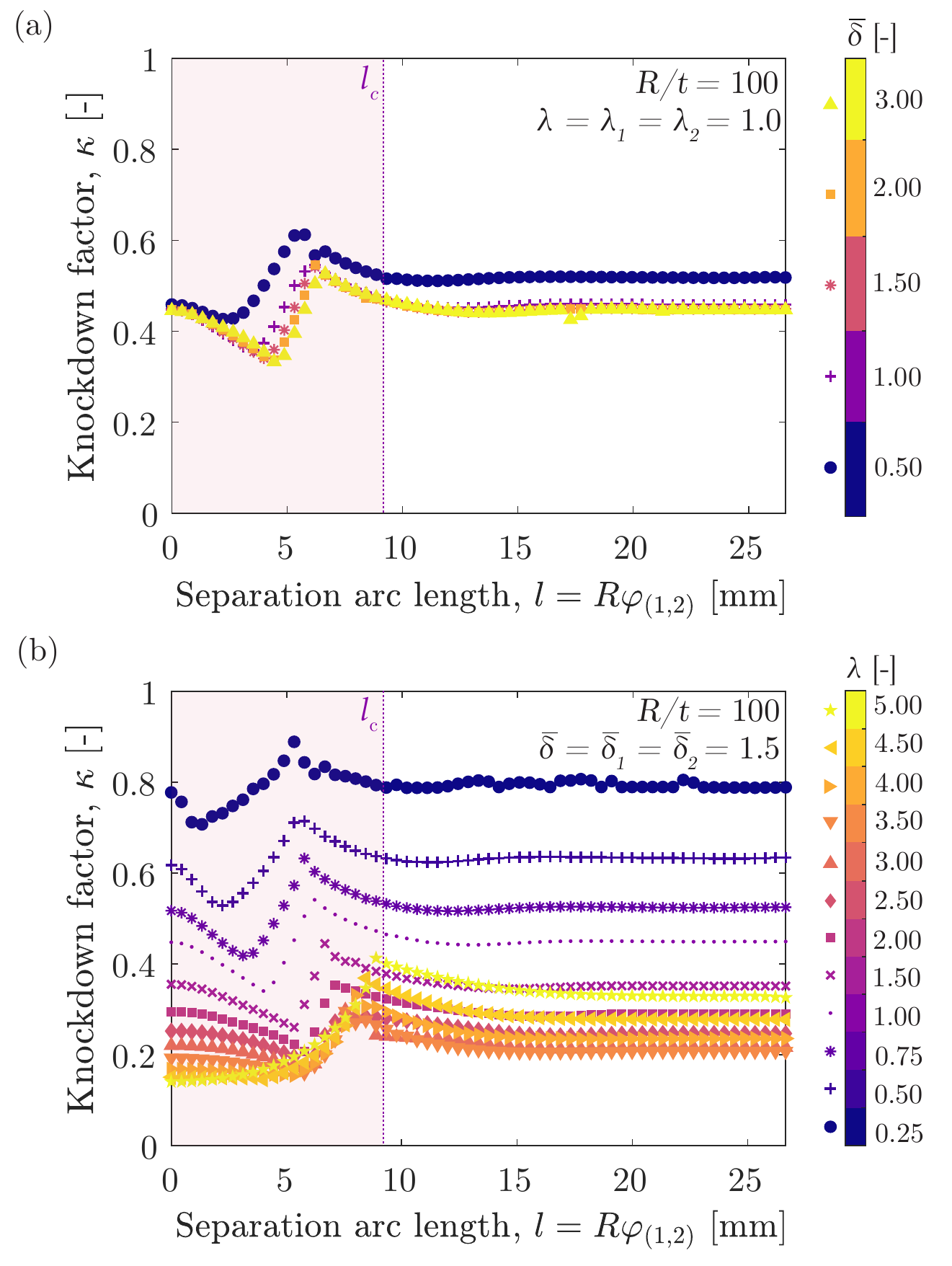}
    \caption{Knockdown factor, $\kappa$, for a shell with $R/t=100$ as a function of the defect-defect arclength, $l$, for identical defects. Panel (a): fixed $\lambda=1$, varying $\overline{\delta} \in [0.5,3]$. Panel (b): fixed $\overline{\delta}=1.5$, varying $\lambda \in [0.25,5]$. Different markers and a color bar distinguish the various parameter values. 
    The vertical dotted line presents the theoretical, critical buckling wavelength, $l_c$ (cf. Eq.~\ref{eq:criticalbucklingwavelength}), for $R/t=100$.}
    \label{fig:lambda_delta}
\end{figure}

In Fig.~\ref{fig:lambda_delta}, we present $\kappa(l)$ curves for a shell with $R/t=100$: in panel (a) for fixed widths ($\lambda=1$) while varying their amplitudes ($\overline{\delta} \in [0.5,3]$), and, in (b), for fixed defect amplitudes ($\overline{\delta}=1.5$) while varying their widths ($\lambda \in [0.25,5]$). In both plots, the vertical lines represent the critical buckling wavelength for a spherical shell, $l_c$, provided in Eq.~\eqref{eq:criticalbucklingwavelength} ~\cite{Hutchinson1967}, for this shell with $R/t=100$. Note that $l_c$ does not depend on any of the defect parameters.  Fig.~\ref{fig:lambda_delta}(a) and Fig.~\ref{fig:lambda_delta}(b) both exhibit non-monotonic $\kappa(l)$, indicative of defect-defect interactions, which consistently occur for $l \lesssim l_c$ (shaded region). For $l\gtrsim l_c$, all curves reach a plateau. Naturally, the specific values of $\kappa_\mathrm{min}$, $\kappa_\mathrm{max}$, and $\kappa_\mathrm{p}$ depend on the actual defect geometry, as extensively investigated in previous studies for single-defect~\cite{lee2016geometric, jimenez_technical_2017, abbasi2023comparing} and  many-defects\cite{derveni2023probabilistic} scenarios.

\begin{figure}[h!]
    \centering
    \includegraphics[width=0.83\columnwidth]{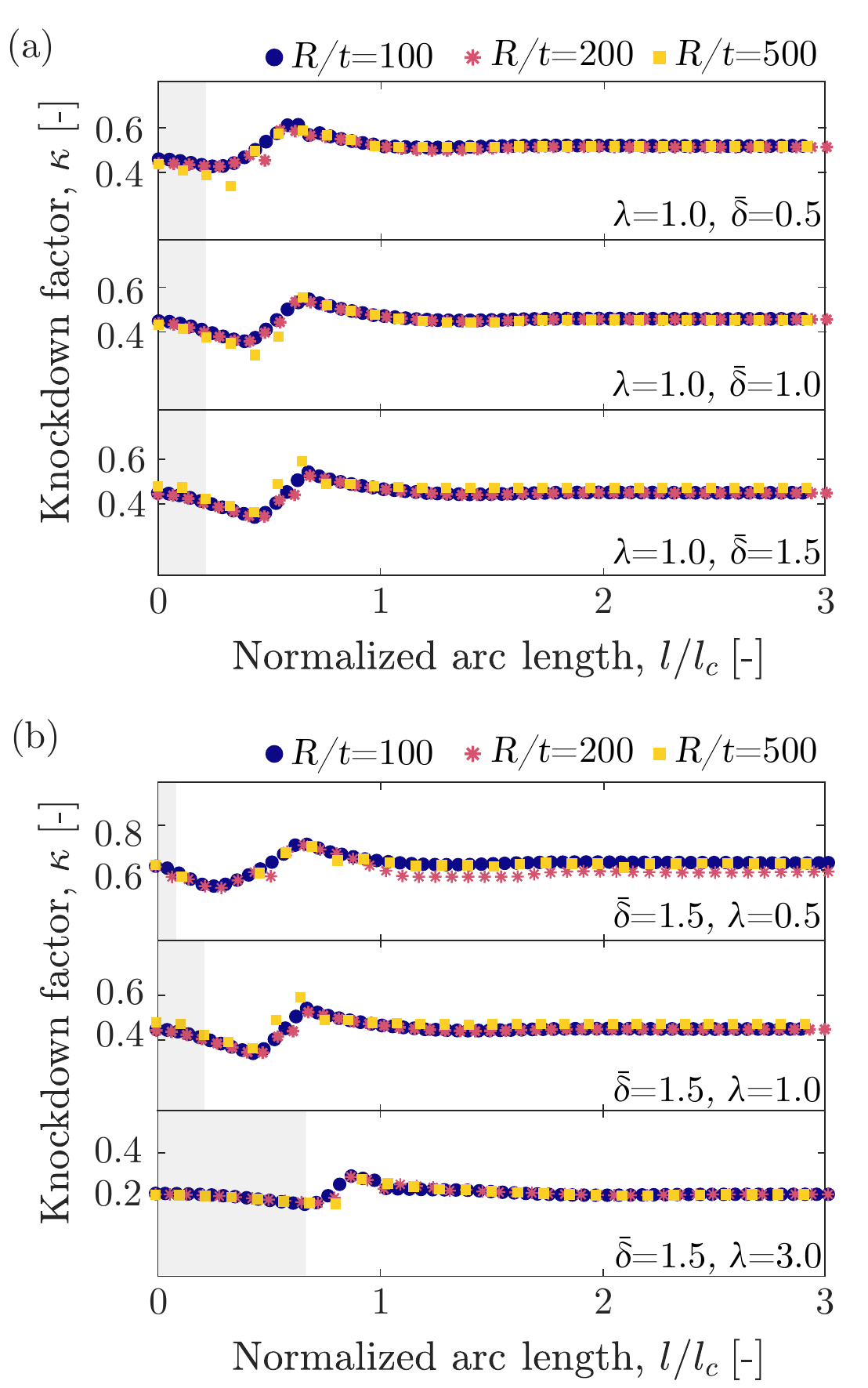}
    \caption{Knockdown factor, $\kappa$, as a function of $l/l_c$, the defect-defect arc length normalized by the critical buckling wavelength defined in Eq.~\eqref{eq:criticalbucklingwavelength}. (a) Constant $\lambda=1$, varying $\overline{\delta}$. (b) Constant $\overline{\delta}=1.5$, varying $\lambda$. The different markers refer to various radius-to-thickness ratios, $R/t$. The shaded areas indicate the regions where the defects overlap, resulting in a single larger defect.}
    \label{fig:collapsed_data}
\end{figure}

We now select some data from Fig.~\ref{fig:lambda_delta}(a), for $\lambda=1$ and $\overline{\delta}=\{0.5,\,1.0,\,1.5\}$, and from Fig.~\ref{fig:lambda_delta}(b), for $\overline{\delta}=1.5$ and $\lambda=\{0.5,\,1.0,\,3.0 \}$, and present them in Fig.~\ref{fig:collapsed_data}(a) and (b) as a function of the normalized arc length $l/l_c$. Additional simulation data for $R/t=200$ and $500$ are included. The shaded regions indicate small angular separations where the two defects overlap (cf. the corresponding 2D profiles in Fig.~\ref{fig:geometry}). It is remarkable that all the $\kappa(l/l_c)$ data collapse, with the emergence of their plateaus past $l/l_c\gtrsim 1$. 

\begin{figure}[h!]
    \centering
    \includegraphics[width=0.9\columnwidth]{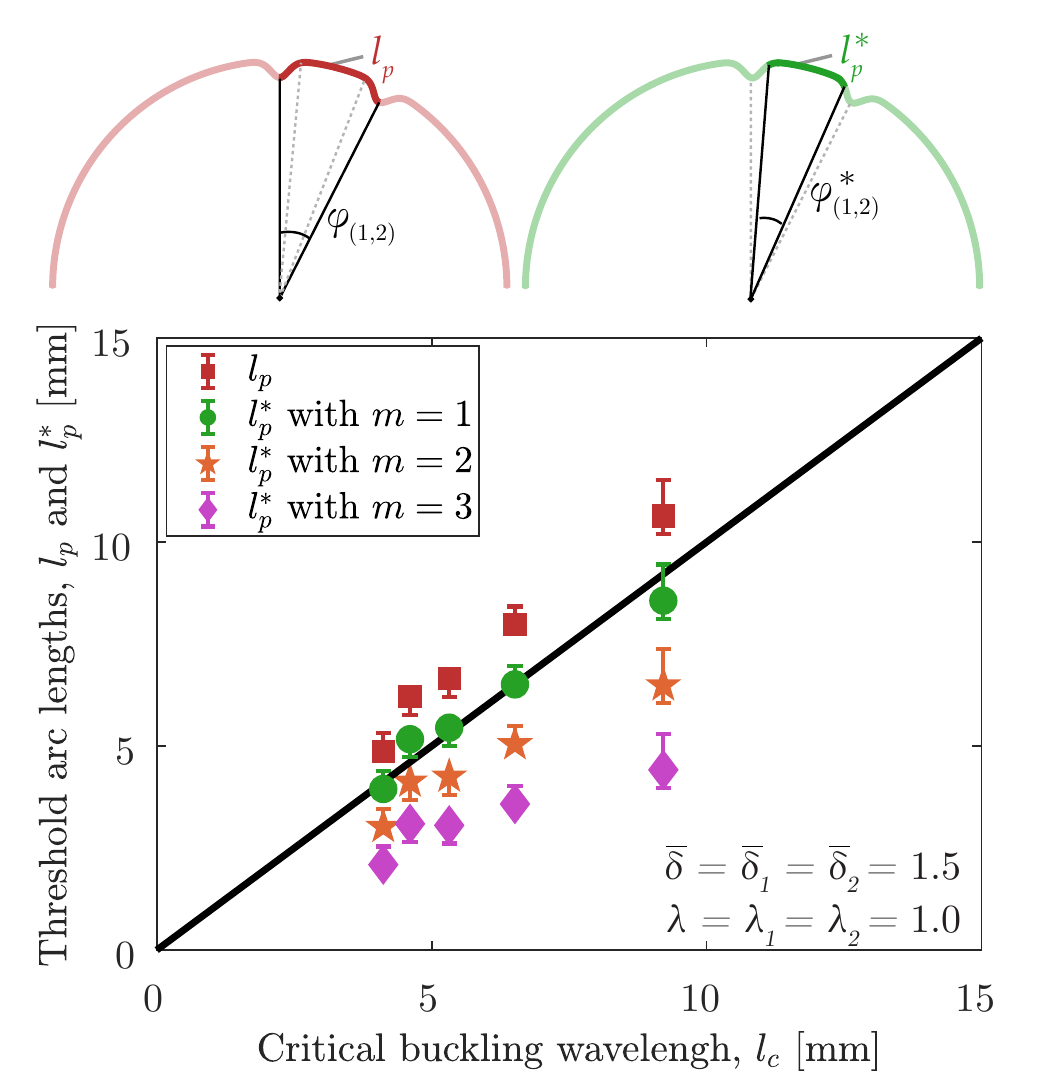}
    \caption{Threshold arc length separations for the interaction regime, $l_p$ and $l_p^*$, versus the critical buckling wavelength, $l_c$,  for identical defects with $\overline{\delta}=1.5$ and $\lambda=1$. Both $l_p=R\varphi_{(1,2)}$ (squares) and $l_p^*=R\varphi_{(1,2)}^*$ (circles for $m=1$, pentagrams for $m=2$, and diamonds for $m=3$) threshold definitions are examined, as illustrated in the 2D schematics (top). The threshold values, $l_p$ and $l_p^*$ are computed as described in the text. The error bars represent $\pm 0.05|\kappa_\mathrm{max}-\kappa_\mathrm{p}|$. The solid line represents $l_\mathrm{p}=l_\mathrm{p}^*=l_c$.}
    \label{fig:threshold}
\end{figure}

The aforementioned observation regarding the onset of the plateau underscores the importance of the critical buckling wavelength, $l_c$, in setting the threshold arc length separation for the defect-defect interaction regime. This finding represents an important step in confirming the hypothesis laid out in Sec.~\ref{sec:hypothesis}. To quantify this threshold, we consider the maximum ($\kappa_\mathrm{max}$) and plateau ($\kappa_\mathrm{p}$) values of the $\kappa(l)$ curves in Figs.~\ref{fig:lambda_delta} and~\ref{fig:collapsed_data}. The threshold separation is defined as the arc length corresponding to the 10\% cut-off: $0.1(\kappa_\mathrm{max}-\kappa_\mathrm{p})$.
An uncertainty of $\pm 0.05(\kappa_\mathrm{max}-\kappa_\mathrm{p})$ is assigned to each threshold value to account for the non-sharp onset of the plateau, consistently with the percentual definitions used in previous work~\cite{jimenez_technical_2017}. 
As mentioned in Sec.~\ref{sec:probdefinition}, there are two possible definitions for the defects separation arc length, $l_p$ or $l^*_p$, depending on whether we consider the center-to-center ($\phi_{(1,2)}$) or the adjusted ($\phi^*_{(1,2)}$) angular separations, respectively. The latter excludes a portion from the core of the defects and was defined in Eq.~(\ref{eq:phi_star_definition}). Schematics illustrating these two definitions are provided in Fig.~\ref{fig:threshold} (top).

At this point, it is important to revisit the Gaussian shape (cf. Eq.~\ref{eqn:geom_gaussiandimple}) of the dimpled imperfections we are considering. Note that, at the local angular coordinate of each defect $\alpha=m\,\alpha_i$, its deviation from the perfect sphere is
$\mathring{w}_i=-\delta_i\,e^{-m}$. Also, $\alpha_i/\sqrt{2}$ can be interpreted as the standard deviation of this Gaussian shape, $\mathring{w}_i(\alpha)$. Therefore, $l_p^*$ can be seen as excluding
some portion of the core of each defect. Taking the values $m=1$, 2, or 3 corresponds to excluding 68.3\%, 95.6\%, and 99.7\% of the defect, respectively \cite{feller1968introduction}. The choice of $m$ determines the extent to which the core of the defect is excluded, with $m=3$ effectively considering the edge-to-edge separation between defects. It is important to note that at $\alpha=\alpha_i/\sqrt{2}$, there is an inflection point in Eq.~(\ref{eq:phi_star_definition}) and $\mathring{w}_i^{\prime\prime}(\alpha_i/\sqrt{2})=0$. 

We have measured $l_p$ or $l^*_p$ as functions of $l_c$, for shells with $R/t\in[100,\,500\}$ and two identical defects with $(\overline{\delta},\,\lambda)=(1.5,\,1.0)$. It is worth noting that the different values of $R/t$ yield different values of $l_c$ according to Eq.~(\ref{eq:criticalbucklingwavelength}); specifically, $l_c$ increases as $R/t$ decreases.
The results shown in Fig.~\ref{fig:threshold} confirm the hypothesis presented in Sec.~\ref{sec:hypothesis}: there is a clear \textit{linear} scaling between  $l_p$ or $l^*_p$ with varying $m$ values (cf. Eq.~\ref{eq:phi_star_definition}) and $l_c$. What is more, when using the $l_p^*$ definition with $m=1$, the data lie on the line $l^*_p=l_c$. This remarkable result demonstrates that the threshold separation for defect-defect interactions is set by the critical buckling wavelength of the shell at the inflection point in the Gaussian profile, $\mathring{w}(\alpha_i)$. Hence, for the remainder of our study, we will adopt the definition of $l_p^*$ with $m=1$.

\begin{figure}[t!]
    \centering
    \includegraphics[width=0.82\columnwidth]{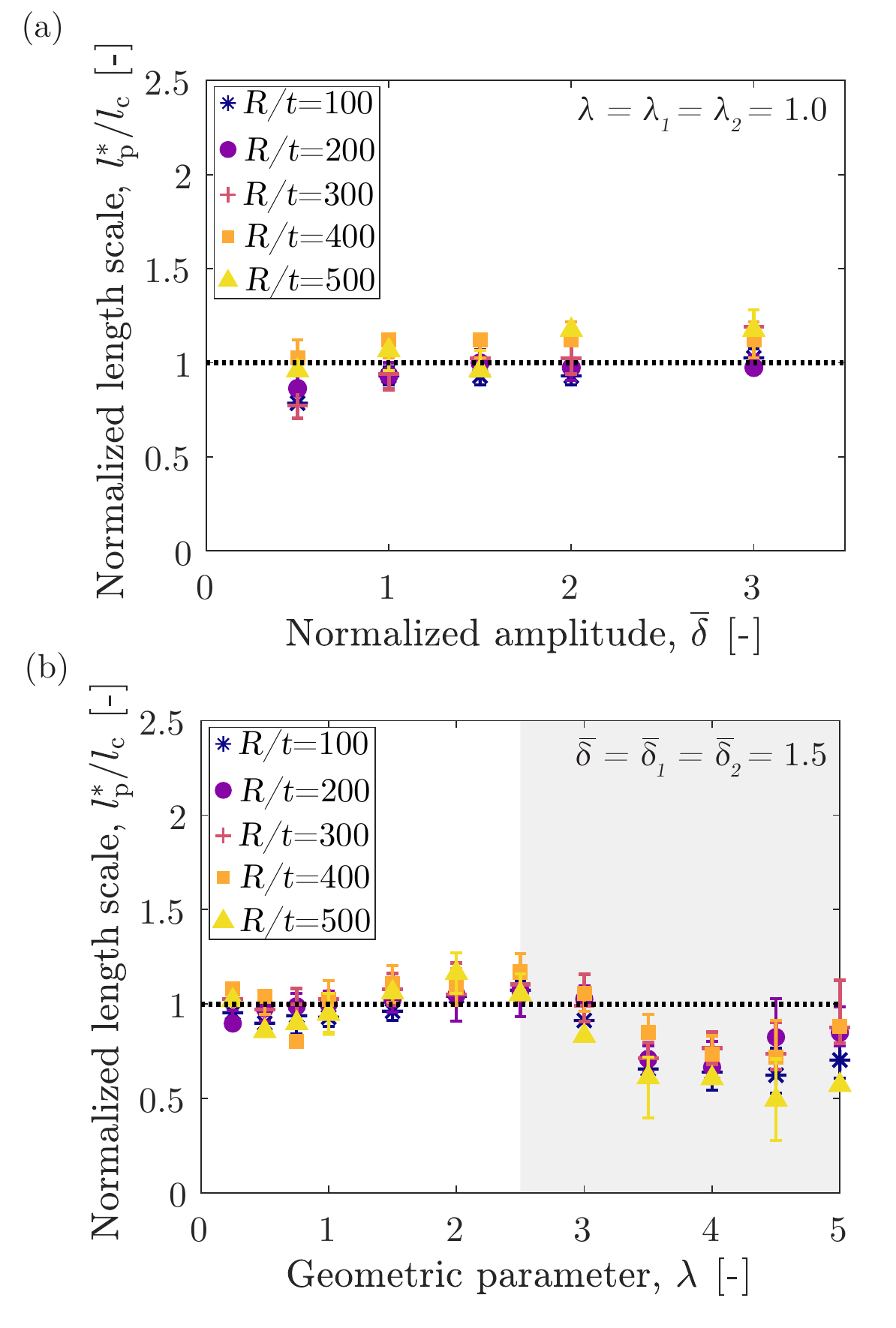}
    \caption{Normalized threshold defect-defect arclength, $l_p^*/l_c$, versus (a) normalized amplitude, $\overline{\delta}$, and (b) normalized width, $\lambda$, for various values of $R/t \in [100,500]$. In panel (a), $\lambda=1$ is kept fixed, and in panel (b), $\overline{\delta}=1.5$ is fixed. Each marker represents a different value of $R/t \in [100,500]$, and the horizontal dashed lines correspond to $l_p^*=l_c$. The shaded area in panel (b) highlights the region where defects tend to overlap, forming a single larger defect.
    }
    \label{fig:threshold_lambda_delta}
\end{figure}

Having examined the specific geometry for an imperfect shell with $(\overline{\delta},\,\lambda)=(1.5,\,1.0)$ (albeit with different $R/t$), we now explore the geometric parameter space more systematically. In Fig.~\ref{fig:threshold_lambda_delta}(a), we plot $l_p^*/l_c$ as a function of $\overline{\delta}$ (with fixed $\lambda=1.0$), and in Fig.~\ref{fig:threshold_lambda_delta}(b) $\lambda$ (with fixed $\overline{\delta}=1.5$), for different $R/t$ values (see legend). Overall, the data consistently aligns closely with $l_\mathrm{p}^*/l_\mathrm{c}=1$ (horizontal dashed line), especially when $\overline{\delta}\geq1$ (Fig.~\ref{fig:threshold_lambda_delta}a) and $\lambda \leq 2.5$ (Fig.~\ref{fig:threshold_lambda_delta}b). In Fig.~\ref{fig:threshold_lambda_delta}(a), $l_\mathrm{p}^*/l_c$ remains approximately constant for all $\overline{\delta} \in [0.5,3]$ and all $R/t \in [100,500]$. 
As also highlighted in Fig.~\ref{fig:lambda_delta}(a), the $l_\mathrm{p}^*/l_c$ data lie almost on top of the dashed line, deviating by at most $20\%$ within the entire range of $\overline{\delta}$ that we explored. More quantitatively, in Fig.~\ref{fig:threshold_lambda_delta}(b), for shells with $\lambda \leq 2.5$, the FEM-measured $l_\mathrm{p}^*$ is in excellent agreement with the analytical result for $l_\mathrm{c}$, within a $16\%$ difference. For wider defects with $\lambda \geq 2.5$, $l_\mathrm{p}^*$ deviates by up to $\approx 50\%$ from $l_\mathrm{c}$. Note that in these shells with wide defects (large $\lambda$ values), the two defects tend to be nearly juxtaposed, as seen in the profiles in Fig.~\ref{fig:geometry}(c) and (d), as well as the shaded region in Fig.~\ref{fig:threshold_lambda_delta}b (for shells with $R/t=100$). We attribute the larger deviations of $l_\mathrm{p}^*/l_\mathrm{c}$ from unity for shells with wide defects to their overlap, which leads to a distorted, imperfect shell geometry. 

\section{Interactions between two different defects}
\label{sec:different}

In the previous section, we examined shells with two identical defects. 
\begin{figure*}[t!]
    \centering
    \includegraphics[width=1.47\columnwidth]{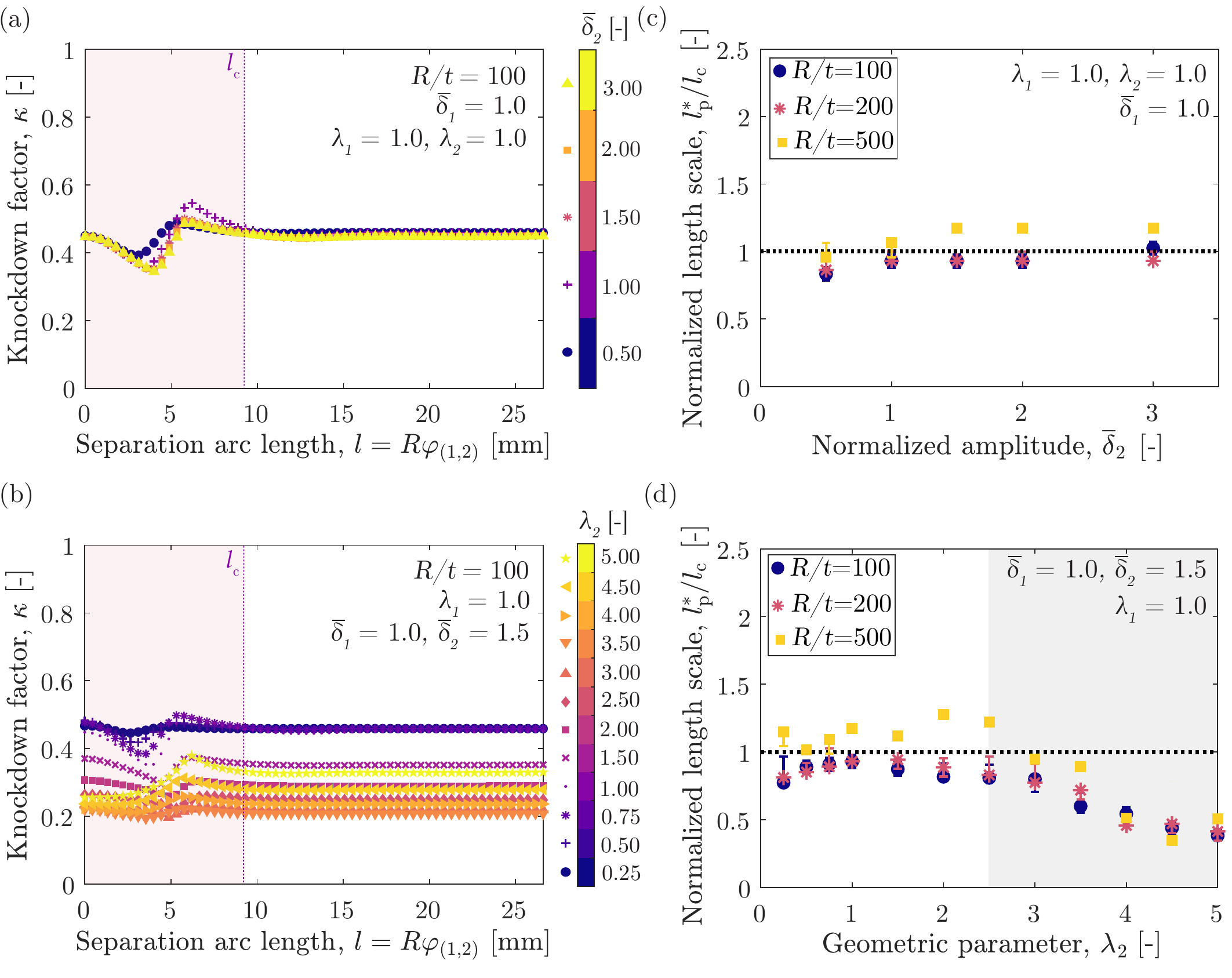}
    \caption{Knockdown factor, $\kappa$, versus arc length, $l$, for shells with $R/t=100$. (a) Fixed $\lambda_1=\lambda_2=1$, $\overline{\delta}_1=1$ and varying $\overline{\delta}_2 \in [0.5, 3]$. (b) Fixed $\overline{\delta}_1=1$, $\overline{\delta}_2=1.5$, $\lambda_1=1$ and varying $\lambda_2 \in [0.25, 5]$. Normalized arclength, $l_p^*/l_c$, versus (c) $\overline{\delta}_2$, and (d) $\lambda_2$ for $R/t \in [100,500]$. Different markers and colors are used to represent different (a) $\overline{\delta}_2$, (b) $\lambda_2$, and (c,d) $R/t$. The vertical dashed lines in panels (a,b) refer to the theoretical prediction of $l_c$ for shells with $R/t=100$, while the horizontal dashed lines in panels (c,d) represent $l_p^*=l_c$. The shaded region in panel (d) indicates the region where defects overlap, forming a single larger defect (shown for $R/t=100$, as a representative example).}
    \label{fig:different_defects}
\end{figure*}
Now, we shift our focus to the case of different defects ($\overline{\delta}_1\neq\overline{\delta}_2$ or $\lambda_1\neq\lambda_2$). We will fix the geometry of the $i=1$ defect at the pole with ($\lambda_{1},\overline{\delta}_1)=(1.0,1.0)$, 
and vary the width ($\lambda_{2}$) and amplitude ($\overline{\delta}_2$) of the second defect.

In Fig.~\ref{fig:different_defects}(a),  we plot the knockdown factor, $\kappa$, as a function of defect-defect arc length separation, $l$, for shells with fixed $R/t=100$ and $\lambda_2=1.0$, while varying $\overline{\delta}_2 \in [0.5,3]$. These $\kappa(l)$ curves are similar to those for the identical-defects case discussed in Sec.~\ref{sec:identical}: $\kappa$ initially decreases to $\kappa_\mathrm{min}$, then increases $\kappa_\mathrm{max}$, before settling to a plateau ($\kappa_\mathrm{p}$). The exact values of $\kappa_\mathrm{min}$, $\kappa_\mathrm{max}$, and $\kappa_\mathrm{p}$ are slightly influenced by the amplitude of the $i=2$ defect, particularly for $\overline{\delta}_2=\{0.5,\,1.0\}$, but not for $\overline{\delta}_2>1.0$, consistent with the known sensitivity of shell buckling to imperfections~\cite{lee2016geometric}.

In Fig.~\ref{fig:different_defects}(b), we present $\kappa(l)$ curves for shells with a fixed $R/t=100$ and $\overline{\delta}_2=1.5$, while  varying $\lambda_2 \in [0.25,5]$. The response of these shells is qualitatively different from the behavior described in the previous paragraph, exhibiting three distinct regimes.  In the first, when $\lambda_2 \leq 1$, the $\kappa(l)$ curves show the same minimum-maximum-plateau dependence described above and in Sec.~\ref{sec:identical}. Since $\lambda_2>\lambda_1$, the plateau
is dictated by the largest ($i=2$) defect. In the second regime, for $1.5 \leq \lambda_2 \leq 3$, the $\kappa(l)$ curves shift, as a whole, to lower values. While a clear minimum is still observed, the maximum becomes less prominent, tending towards $\kappa_\mathrm{max} \to \kappa_\mathrm{p}$. In this regime, the buckling is still dictated by the largest $i=2$ defect. In the third regime, for $\lambda \geq 3.5$, the $\kappa(l)$ curves shift upwards. 

In Fig.~\ref{fig:different_defects}(a,b), the vertical dotted lines represent the critical buckling wavelength, $l_c$, defined in Eq.~(\ref{eq:criticalbucklingwavelength}), with $R/t=100$. Similarly to the case of identical defects, we observe that the region (shaded) of interaction for these shells with two different defects lies within $l<l_c$. As in Sec.~\ref{sec:identical}, we also compute the normalized threshold for defect-defect interactions (onset of the plateau of the $\kappa(l)$ curves), $l_\mathrm{p}^*/l_\mathrm{c}$, for the present case of different defects. These results are presented in Fig.~\ref{fig:different_defects}(c,d). 

In Fig.~\ref{fig:different_defects}(c), when fixing $\overline{\delta}_1$, $\lambda_1$, and $\lambda_2$ , we observe that 
$l_\mathrm{p}^*/l_\mathrm{c}\approx 1$ (within $17\%$) across the whole range of $\overline{\delta}_2$. This finding reinforces that $\overline{\delta}$ is not critical in determining the onset of defect interactions, consistently with the identical-defects case (Fig.~\ref{fig:threshold_lambda_delta}a). The behavior becomes less straightforward when varying $\lambda_2$ while fixing $\overline{\delta}_1$, $\lambda_1$, and $\overline{\delta}_2$, (see Fig.~\ref{fig:different_defects}d). Here, $l_p^*/l_c$ remains near unity for $\lambda_2 \leq 3$, with a deviation of around $22\%$ for $\lambda_2 \in[0.25,\,1]$ and $28\%$ for $\lambda_2 \in[1.5,\,3]$. However, when $\lambda_2 \geq 3.5$, $l_p^*/l_c$ progressively drops below unity, reaching approximately $0.4$. Recalling the profiles in Fig.~\ref{fig:geometry}(d), we note that the edges of the narrow $i=1$ defect overlap with the wider $i=2$ defect for larger values of $\lambda_2$. Thus, the shell geometry deviates substantially from a perfect sphere, and the critical buckling wavelength in Eq.~(\ref{eq:criticalbucklingwavelength}) no longer sets the edge of the interaction region. This complex behavior, arising from the increasing overlap of the defects and the nontrivial shell geometries, falls beyond the scope of the present work and warrants further investigation.

Note that, in Fig.~\ref{fig:different_defects}(c,d), while $l_p^*/l_c$ remains close to unity for intermediate values of $\lambda_2$, the thinnest shells with $R/t=500$ exhibit notable discrepancies compared to the $R/t=\{100,\,200\}$ shells (the results for these two are almost overlapping). We have conducted comprehensive mesh-convergence tests, and it appears that the discrepancies are not due to the discretization. Instead, we attribute these deviations to the higher fluctuations observed in the measured $\kappa(l)$ curves, especially in the plateau region, which in turn affects the measurement of $l_p^*$ using the 10\% criterion introduced in Sec.~\ref{sec:identical}.

\section{Conclusions}
\label{sec:conclusions}

Using experimentally validated FEM simulations, we investigated the effect of defect-defect interactions on the buckling of pressurized hemispherical shells containing two dimpled imperfections. We examined cases of identical and different defects, varying their geometric parameters (amplitude, $\overline{\delta}_i$, and width, $\lambda_i$) and their relative separation. We measured the knockdown factor (the normalized critical buckling pressure), $\kappa$, for these imperfect shells as a function of the angular separation, $\varphi_{(1,2)}$, between their two defects. We then used $\varphi_{(1,2)}$ to define an arc length separation $l=R\varphi_{(1,2)}$. Our findings revealed significant defect-defect interactions when the two defects are in close proximity, leading to non-monotonic behavior in $\kappa(l)$, below a threshold in $l$. We modified the definition of this interaction threshold, denoted as $l_p^*$, which corresponds to the inflection point of the Gaussian profile. Beyond $l^*_p$, the  $\kappa(l)$ curves reached a plateau, indicating diminished interactions and the dominance of the largest defect in dictating the knockdown factor.

The main contribution of our study lies in establishing that the onset of defect-defect interactions is determined by the critical buckling wavelength~\cite{Hutchinson1967}, as $l_p^*\approx l_c$ (cf. Eq.~\ref{eq:criticalbucklingwavelength}). This result is valid for defects with $\lambda_i<3$, regardless of whether they are identical or different. However, for wider defects, the dimples tend to overlap, and the shell geometry becomes increasingly distorted. The defect amplitude, $\overline{\delta}_i$, plays a negligible role in setting $l_p^*$. It is important to note that $l_c$ depends only on the radius, $R$, and thickness, $t$, of the shell (other than the Poisson ratio, which was fixed to $\nu=0.5$ throughout our study).

We hope that our results will stimulate further interest in harnessing defect-defect interactions to enhance the buckling response of spherical shells or inspire the development of novel functional mechanisms derived from these interactions.

 \vspace*{5mm}
\begin{acknowledgments}
We are grateful to John Hutchinson for insightful discussions, which inspired the scope and findings of our study. A comment by him on our previous work~\cite{derveni2023probabilistic} was at the source of the hypothesis that $l_c$ dictates the onset of defect-defect interactions. We also thank Michael Gomez for his invaluable feedback on the results presented in this manuscript.
\end{acknowledgments}

  \vspace*{5mm}
\noindent \textbf{Disclosure on the usage of large language model (LLM)}

We used the Large Language Model (LLM) -- ChatGPT (GPT-4 architecture, May 12 Version) -- in the drafting of this manuscript for grammar and language refinement. We only employed the following two prompts: ``\texttt{fix grammar and typos}" and ``\texttt{provide alternative phrasing for}." Nonetheless, all final decisions and content in the manuscript were made and thoroughly reviewed by the authors. As supplementary information, we have included a commented "\LaTeX diff" version that compares the nearly final draft prior to using ChatGPT with the present final version, noting that the latter also includes several minor edits made by the authors \textit{a posteriori}.
\bibliographystyle{apssamp}
\bibliography{apssamp}

\end{document}